# DeepSIBA: Chemical Structure-based Inference of Biological Alterations

C.Fotis[1+], N.Meimetis[1+], A.Sardis[1], LG. Alexopoulos[1*]

1 BioSys Lab, National Technical University of Athens, Athens, Greece

+ Equal contributions

∗ Correspondence to: leo@mail.ntua.gr

## Abstract

Predicting whether a chemical structure shares a desired biological effect can have a significant impact for in-silico compound screening in early drug discovery. In this study, we developed a deep learning model where compound structures are represented as graphs and then linked to their biological footprint. To make this complex problem computationally tractable, compound differences were mapped to biological effect alterations using Siamese Graph Convolutional Neural Networks. The proposed model was able to learn new representations from chemical structures and identify structurally dissimilar compounds that affect similar biological processes with high precision. Additionally, by utilizing deep ensembles to estimate uncertainty, we were able to provide reliable and accurate predictions for chemical structures that are very different from the ones used during training. Finally, we present a novel inference approach, where the trained models are used to estimate the signaling pathways affected by a compound perturbation in a specific cell line, using only its chemical structure as input. As a use case, this approach was used to infer signaling pathways affected by FDA-approved anticancer drugs.

# 1 Introduction

Early stage drug discovery aims to identify the right compound for the right target, for the right disease. A very important step in this process is hit identification, in which compounds that exhibit strong binding affinity to the target protein are prioritized. Traditionally, the most widely employed method for *in-vitro* hit identification is High Throughput Screening (HTS). *In-vitro* HTS can produce hits with strong binding affinity that may later be developed into lead compounds through lead optimization. However, due to the vast chemical space, even large scale *in-vitro* HTS offers limited chemical coverage and does not guarantee the biological efficacy and low toxicity of the identified hits. On this front, several *in-silico* methods, either Computer Aided Drug Design (CADD) or systems pharmacology-based, aim to assess the structural and biological effects of compounds with the optimal goal of improving the success rate of identified hits.

The development of Computer Aided Drug Design (CADD) methods has enabled the virtual High Throughput Screening (vHTS) of vast compound libraries, thus effectively increasing the search space of hit identification. CADD methods for vHTS prioritize compounds, which are likely to have activity against the target, for further experiments and are broadly categorized into structure-based and ligand-based.[1] Structure-based CADD approaches require the solved 3D structure of the target protein, either through X-ray crystallography or NMR spectroscopy and focus on docking simulations to assess protein-ligand complexes. On the other hand, ligand-based virtual screening is used when the 3D structure of the target is unknown and involves the calculation of 2D or 3D structural similarities between a known active ligand and a virtual library. Structural similarity screening is based on the hypothesis that similar chemical structures will cause similar response.[2] However, there are many cases of compounds and drugs, which although structurally dissimilar, have the same biological effect, either because of off-target effects or by targeting proteins in the same pathway.[3] As a whole, vHTS approaches focus on optimal binding affinity rather than on the system effects of compounds, which are more closely related to clinical efficacy and toxicity.[4]

Advances in systems-based approaches and 'omics technologies have led to the development of systems pharmacology methods that aim to lower the attrition rates of early stage drug discovery. Systems pharmacology approaches couple 'omics data with knowledge bases of molecular interactions and network analysis methods in order to assess compounds based on their biological effect.[5] One approach that has gained considerable attraction is the use of gene expression (GEx) profiling to characterize the systematic effects of compounds. On this front, Verbist *et al*. showed how GEx data were able to influence decision making in eight drug discovery projects by uncovering potential adverse effects of the lead compounds.[6] Additionally, Iorio *et al*. utilized similarities between drugs' transcriptional responses to create a drug network and identified the mechanism of action of new drugs based on their position in the network.[7] Since its release, the Connectivity Map (CMap) and the LINCS project have been a cornerstone of transcriptomic-based approaches by providing a large scale database of transcriptomic signatures from compound perturbations along with essential signature matching algorithms.[8,9] CMap's approach is based on the hypothesis that compounds with similar transcriptomic signatures will cause similar physiological effects on the cell and has been widely adopted by the field of drug repurposing.[10] However, signature-based approaches are not only limited in the search space of compounds with available GEx data but are also missing key chemical information that is pivotal for drug design. A computational framework that holds promise in connecting the chemical structure with the biological effect of compound perturbations is machine learning.

Traditionally, machine learning methods like Support Vector Machines (SVM), random forests (RF) and neural networks (NN) have been successfully employed in drug discovery applications, including property prediction and Quantitative Structure Activity Relationship (QSAR) modeling.[11] The recent increase in available data and computing power has given rise to Deep Learning (DL) methods for various tasks, including bioactivity and toxicity prediction as well as de-novo molecular design.[12-14] In this regard, the DeepChem library along with MoleculeNet have been a cornerstone of DL approaches in drug discovery by providing a plethora of architectures along with benchmark datasets for their comparison.[15, 16] DL methods offer the advantage of flexible end-to-end architectures that learn task specific representations of chemical structures, without the need for precomputed features.[17] One particular DL architecture that has achieved state of the art results in several drug discovery benchmark datasets is the Graph Convolutional Neural Network (GCNN).[18] Molecular GCNNs operate on chemical structures represented as undirected graphs, with nodes being the atoms and edges the bonds between them. They are an extension of Convolutional Neural Networks (CNN) to non-Euclidean data and aim to learn neighborhood level representations of the input graph. Torng *et al.* implemented GCNNs to encode both compounds and protein binding pockets for the task of protein-ligand binding prediction, while also utilizing unsupervised graph convolutional autoencoders for model pretraining.[19] Kearnes *et al.* developed the Weave graph convolution module, which encodes both atom and bond representations and combines them using fuzzy histograms to extract meaningful molecule-level representations.[20] The Weave module was applied to ligand-based virtual screening and showed improved performance compared to methods using precomputed features. Despite their improved performance over traditional ML methods, end-to-end models including GCNNs are still prone to generalization errors on new chemical scaffolds. This is mainly because of the limited coverage of the chemical space by the training data.[21] In order to tackle this limited chemical coverage, methods like one-shot learning are promising candidates for drug discovery applications. One-shot learning techniques, such as Siamese and Matching networks, aim to learn a meaningful distance function between related inputs and have shown increased performance over traditional methods in tasks with few data points.[22-25] Altae-Tran *et al.* implemented one-shot learning for drug discovery by combining graph convolutions and Long Short Term Memory (LSTM) networks with attention and achieved better results than traditional GCNNs.[26] Furthermore, for drug discovery applications, uncertainty estimation is crucial, since incorrect predictions e.g. regarding toxicity can lead to incorrect prioritization of compounds for further experimental testing. Methods for quantifying predictive uncertainty in deep learning applications include test-time Dropout, deep ensembles and Bayesian NNs.[27-30] Ryu *et al.* developed Bayesian GCNNs for molecular property, bioactivity and toxicity predictions and showed that quantifying predictive uncertainty can lead to more accurate virtual screening results.[31] The flexibility provided by end-to-end architectures along with the aforementioned methods for low data tasks can provide a flexible framework to incorporate both systems and ligand-based approaches for early stage drug discovery.

In this paper, we employ deep learning to decipher the complex relationship between a compound's chemical structure and its biological effect. To make this complex problem computationally tractable, we focus on learning a combined representation and distance function that maps structural differences to biological effect alterations. For this task, we propose a deep Siamese GCNN model called deepSIBA. DeepSIBA takes as input pairs of compound structures, represented as graphs and outputs their biological effect distance, in terms of enriched biological processes (BPs) along with an estimated uncertainty. In order to account for the biological factors that influence the learning task, we train cell line-specific deep ensembles only on carefully selected chemical structures, for which high quality GEx data are available.

The performance of our approach was evaluated with a realistic drug discovery scenario in mind, as a parallel to vHTS, where gene expression data are available for only one compound per pair and compared with a recently proposed architecture for a similar task.[32] Finally, we present a novel KNN-like approach, in which the trained models can be used to infer the signaling pathway signature of a target compound, without available GEx data. As a use case, this approach was tasked to infer the signaling pathway signature of approved anticancer drugs for which no transcriptomic signatures are available in our data sets, using only their chemical structure as input. DeepSIBA can be used in combination with existing pipelines for vHTS to identify structures that not only exhibit maximal binding affinity but also cause a desired biological effect. By incorporating this systems-based approach in vHTS, the hit identification process can produce candidates with improved clinical efficacy and toxicity.

## 2 Material and methods

### 2.1 The deepSIBA approach

The overview of our approach is presented in Figure 1. Transcriptomic signatures from compound perturbations along with their respective chemical structures were retrieved from the CMap dataset.[9] For each compound perturbation, normalized enrichment scores (NES) of GO terms related to BPs were calculated using Gene Set Enrichment Analysis (GSEA). Afterwards, the lists of enriched BPs were ranked based on NES and a Kolmogorov-Smirnov based distance function, similar to GSEA, was used to calculate their pairwise distance (Figure 1A). During the learning phase, the proposed model is trained to predict the pairwise distance between compounds' affected BPs using only their chemical structure as input. The input chemical structures are represented as undirected graphs, with nodes being the atoms and edges the bonds between them and encoded using a Siamese GCNN architecture (Figure 1B). In our approach, compounds with available GEx data, representing a small portion of the chemical space, serve as reference for the inference phase. During inference, the model is tasked to predict the biological effect distance between reference and unknown compounds (without available GEx data).

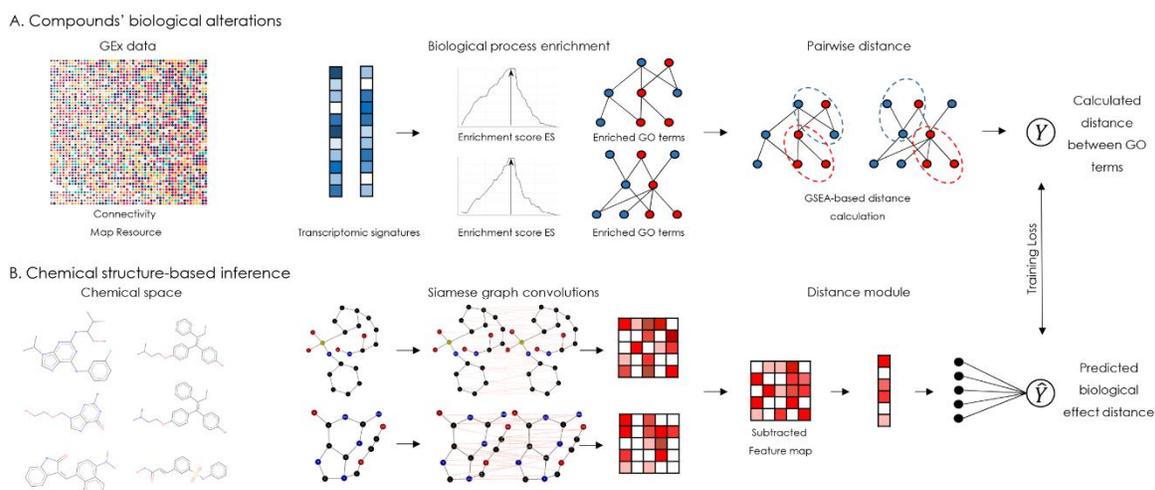

Figure 1 Schematic overview of deepSIBA. (A) Pairs of transcriptomic signatures following compound treatment are retrieved and enriched GO terms for BPs are calculated. The pairwise distance between enriched BPs is calculated using a Kolmogorov-Smirnov based function ($Y$). (B) Pairs of chemical structures are represented as molecular graphs and encoded by a deep learning model using Siamese graph convolutions. Compounds' feature maps are then subtracted and a score, which represents their distance between enriched BPs, is predicted ($\hat{Y}$). The deep learning model is trained by minimizing the loss between predicted ($\hat{Y}$) and calculated distance ($Y$).

## 2.2 Data preprocessing and quality control

Transcriptomic signatures (level 5 z-score transformed) following compound treatment were downloaded from the L1000 CMap resource.[33] In this project, only the differential expression of the 978 landmark genes in the L1000 assay was considered. For each signature, a quality score was derived, based on its transcriptional activity score (TAS), the number of biological replicates and whether the signature is considered an exemplar. This quality score ranges from Q1 to Q8, with Q1 representing the highest quality. TAS is a metric that measures a signature's strength and reproducibility and is calculated as the geometric mean of the number of differentially expressed (DEx) transcripts and the 75$^{th}$ quantile of pairwise replicate correlations. Furthermore, exemplar signatures are specifically designated for further analysis in the CLUE platform.[34] For each compound per cell line, among signatures from different dosages and time points, the signature with the highest quality was selected. An overview of the processed dataset is presented in Electronic Supplementary Information (ESI) 1.1.

## 2.3 Biological process enrichment and pairwise distance calculation

Gene Ontology (GO) terms for biological processes (BP) involving the landmark genes of the L1000 assay were retrieved using the topGO R package in Bioconductor.[35] Only GO terms with at least 10 genes were considered. For each signature, GO term enrichment was calculated using the R package FGSEA in Bioconductor.[36] Thus, the gene-level feature vector of each perturbation was transformed to a BP-level feature vector of Normalized Enrichment Scores (NES). Pairwise distances between BP-level feature vectors were calculated similar to Iorio *et al.*[7], using the R package Gene Expression Signature in Bioconductor.[37] Given two feature vectors ranked by NES, A and B, GSEA is used to calculate the ES of the top and bottom GO terms of A in B and vice versa. The distance between the vectors is computed as $1 - \frac{ES_{A\ in\ B} + ES_{B\ in\ A}}{2}$ and ranges from 0 to 2. An important parameter that can introduce bias in the distance calculation is the number of top and bottom GO terms to consider during GSEA. On this front, an ensemble approach was developed, by calculating pairwise distances between BP-level feature vectors for 5 different numbers of top and bottom GO terms. The numbers we considered were selected based on the average number of significantly enriched GO terms across all perturbations in the dataset (see ESI 1.2 for details). The distance scores were finally averaged and normalized between 0 and 1.

## 2.4 Siamese GCNN architecture

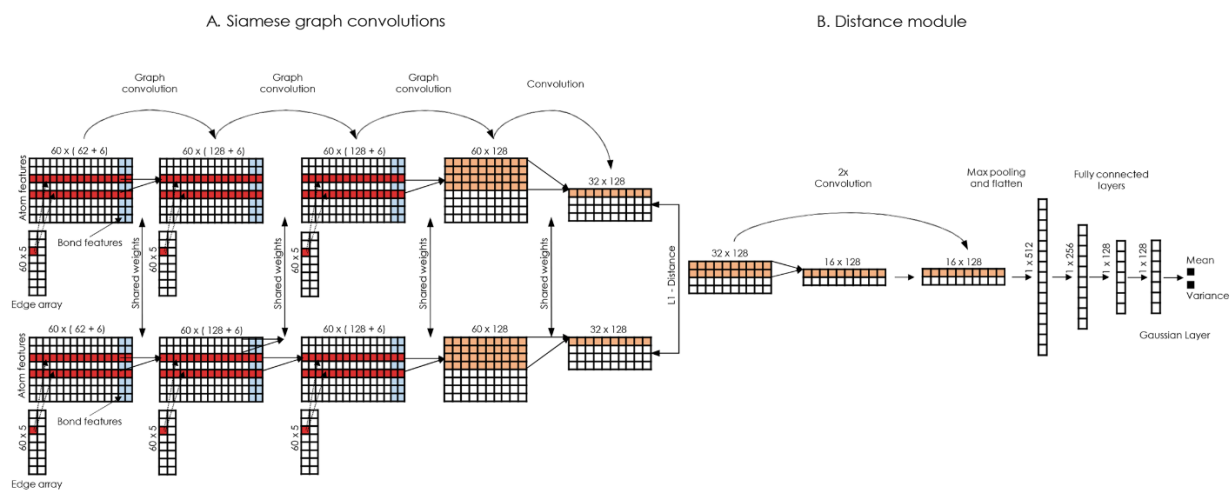

Figure 2 Schematic representation of the model's architecture. (A) Siamese graph convolutional encoders; compounds' molecular graphs are encoded using 2 encoders with shared weights (Siamese). Each encoder consists of 3 graph convolution and 1 convolution layers. (B) Architecture of the distance module; the distance module consists of 2 convolution, 3 fully connected and 1 Gaussian regression layers.

A schematic representation of our model's architecture is presented in Figure 2. The learning model takes as input the chemical structures of compound pairs and predicts their biological distance, at the level of affected biological processes (GO terms). Regarding the input, chemical structures are represented as undirected graphs, with nodes being the atoms and edges the bonds between them. Each input is encoded using 3 matrices: the atom array, which contains atom-level features, the bond array, which contains bond-level features and the edge array, which describes the connectivity of the compound (see ESI 2.1 for details). The learning model consists of two Siamese encoders (shared weights) that embed the input graphs into a high dimensional latent space and a trainable distance module that outputs the final distance prediction. The Siamese encoders consist of 3 graph convolutional layers that learn neighborhood-level representations, followed by a convolutional layer that extracts compound-level features (Figure 2A). Graph convolutions were implemented similar to Duvenaud *et al.*[18] (see ESI 2.2 for details). The overall goal of the Siamese encoder is to learn task-specific compound representations. The feature maps of the last Siamese layer are then subtracted and their absolute difference is passed to the distance module. The distance module consists of 2 convolutional layers, which extract important features from the difference of the feature maps and 3 fully connected layers that aim to combine those features, while progressively reducing the dimensions (Figure 2B). Finally, a Gaussian regression layer outputs a mean and variance of the biological effect distance between the compound pair. By treating the distance as a sample from a Gaussian distribution with the predicted mean and variance, the model is trained end-to-end by minimizing the negative log-likelihood criterion[28] given by

$$-log p_\theta(y_n|X_n) = -\frac{1}{2} log \sigma_\theta^2(x) - \frac{1}{2\sigma_\theta^2(x)}\left(y - \mu_\theta(x)\right)^2 + constant.$$

For each cell line, an ensemble model combining 50 models was created. The ensemble's output is also a Gaussian, with mean and variance calculated from the uniformly weighted mixture of each model. The coefficient of variation (CV) of the Gaussian mixture is used as the model's estimate of predictive uncertainty. The model's hyperparameters, along with the equations for the Gaussian mixture's mean and variance are presented in ESI 2.3 and 2.4.

## 2.5 Dataset splitting and evaluation metrics

For each cell line, available compounds were split into training and test. Each cell line specific training set consists of the pairwise distances between training compounds' affected BPs, while each test set contains distances between test and training compounds. Additionally, the Tanimoto similarity between the ECFP4 fingerprints of all training and test compounds was calculated and test compounds that exhibited a similarity higher than 0.85 to any training compound were excluded. An overview of the training and test sets is presented in ESI 3.1. Across all test scenarios, model performance was evaluated in terms of Mean Squared Error (MSE), Pearson's r and precision. Even though the learning task is a regression problem, given its nature and potential applications, precision is important in order to avoid false positive hits for validation experiments. The distance threshold to calculate precision was set at 0.2, based on the distance distribution of duplicate compound signatures. Duplicate signatures indicate transcriptomic signatures from the same compound perturbation, cell line, dose and time point that were assayed on different L1000 plates. A thorough investigation of the distance threshold to distinguish compounds with similar biological effect is presented in ESI 4.1.

## 2.6 Signaling pathway inference for target structure

The predictions of a trained deepSIBA model can be used to infer a pathway signature for a target structure without the need for GEx data, in terms of the most upregulated and downregulated signaling pathways. The inference approach is similar to the k-Nearest Neighbor algorithm (KNN). Given a target structure, a trained ensemble model for the cell line of choice is used to predict all pairwise distances between target and training compounds. The predicted distance represents the difference between compounds' enriched BPs (GO terms). Training set compounds with predicted distance less than a specified threshold $d_{th}$ are selected as the target's neighbors. If a target structure has more than $k$ neighbors, a signaling pathway signature can be inferred in the following way. For each neighbor $N_i$, the lists of the top 10 most upregulated and most downregulated pathways, based on NES, are constructed. Pathway enrichment is calculated using FGSEA with KEGG as a knowledge base.[38] KEGG signaling pathways were chosen for inference due to their interpretability. Signaling pathways that appear in the neighbors' lists with a frequency score higher than a threshold $f_{th}$ are selected. Additionally, to account for signaling pathways that are frequently upregulated or downregulated in the set of training compounds, a p-value for each inferred pathway is also calculated. On this front, sets of $k$ neighbors are randomly sampled 5000 times from the training set and a Null distribution of frequency scores for each pathway is derived. A p-value is computed as the sum of the probabilities of observing equally high or higher frequency scores. Finally, only pathways with p-value lower than a threshold $p_{th}$ are inferred. Thus, for each chemical structure, our approach infers two signatures of variable length (up to 10 each) of potentially downregulated and upregulated pathways respectively. For the MCF7 cell line, the aforementioned thresholds and parameters of the inference approach were selected by evaluating the results, in terms of precision and number of inferred pathways, on its respective test set (see ESI 5.1 for details).

# 3 Results and discussion

## 3.1 Biological factors influence the model's learning task

The presented model is tasked to predict the biological effect distance between compounds, using their molecular graphs as input. Considering that this distance is calculated from experimental GEx data following compound treatment, there are specific biological factors that can influence the learning task. The CMap dataset contains over 110K transcriptomic signatures from over 20K compounds assayed across 70 cell lines. By carefully analyzing these signatures and their pairwise distances, we were able to pinpoint the most influential factors and identify their effect on the model's target value.

**3.1.1 The variation in quality of GEx data is reflected on the calculated distance value.** The quality of gene expression data, from which transcriptomic signatures in the Connectivity map were derived, varies across compound perturbations. In our case, this variation in data quality is especially important. On this front, a categorical quality score, ranging from Q1 to Q8, was assigned to each signature, with a score of Q1 representing the highest quality (see ESI 1.1). In order to assess the effect of signature quality, distributions of distances between duplicate transcriptomic signatures for different quality scores were examined and are presented in Figure 3A. As expected, Q1 duplicate signatures are very similar and their distances are centered near a small value. However, this is not the case for Q2 duplicate signatures, where differences in differentially expressed genes are prominent even when all the perturbation parameters (compound, cell line, dose, time) are kept constant (red line in Figure 3A). It is clear that signature quality significantly affects the distribution of the model's target variable.

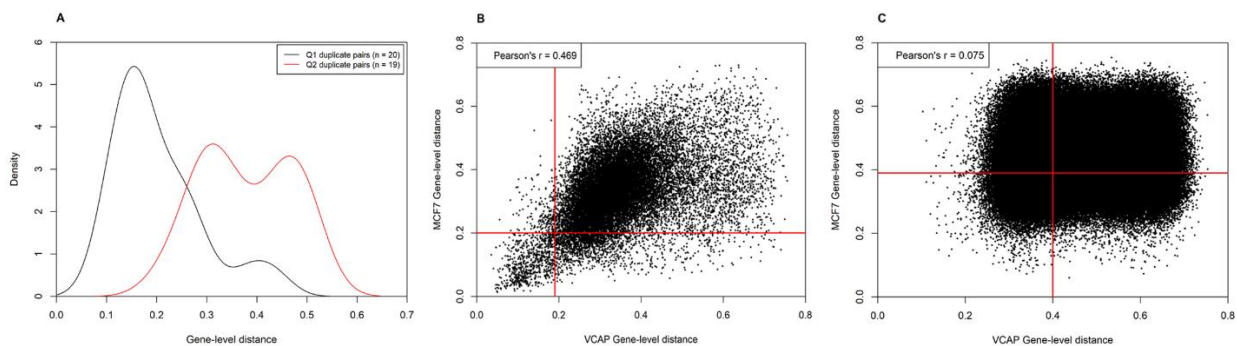

Figure 3 Quality and cell line effect on the calculated distance between compounds' transcriptomic signatures. (A) Distribution of distance values between Q1 duplicate compound signatures (black line) and Q2 duplicate compound signatures (red line), for the MCF7 cell line. (B) Scatterplot of distances between Q1 transcriptomic signatures for the same compound pairs in the MCF7 and VCAP cell lines. Each point in the plot represents a pair of compounds with available transcriptomic signatures in both cell lines. The red lines, at 0.2 for MCF7 and 0.19 for VCAP indicate the mean + standard deviation of the distribution of distances between Q1 duplicate signatures for each respective cell line; (C) similar scatterplot to (B) but for Q2 transcriptomic signatures.

**3.1.2 Distances between transcriptomic signatures vary across cell lines.** Compound response, in terms of DEx genes, is highly dependent on the cellular model. Due to different genetic backgrounds and gene expression patterns the same compound perturbation will have different transcriptomic signatures between cell lines.[39] This dependence, directly affects the distance between compounds' transcriptomic signatures across cell lines. The relationship between gene-level distances of the same compound pairs calculated in the MCF7 and VCAP cell lines, for Q1 and Q2 signatures, is shown in Figures 3B and 3C. In general, Q1 transcriptomic distances of the same compound pair in the 2 examined cell lines are moderately correlated (Pearson's r = 0.468). However, there is a significant number of compound pairs which have similar transcriptomic signatures in one cell line but not in the other (lower right and upper right quadrants of Figure 3B). Such cases are more prominent for compound pairs with Q2 signatures (lower right and upper left quadrants of Figure 3C). Thus, the cell line effect poses a problem for the proposed learning task by providing a one-to-many mapping between input (pair of chemical structures) and output (distance between signatures). One common approach to address this issue is to aggregate transcriptomic signatures or distances across cell lines. While aggregating enables the training of a general model on all available compound pairs, it can often produce misleading results and cause information loss.

**3.1.3 Compounds' biological effects are better represented on a functional level.** A distance function that operates directly on transcriptomic signatures does not account for smaller differences in the DEx of genes that belong to the same biological pathway. Thus, the similar effect between perturbations, in terms of enriched BPs, might not be clearly reflected on their gene-level distance. On this front, a comparison of BP and gene-level distances between cell lines for the knockdown of the MYC gene with shRNA is presented in Figure 4. MYC is an oncogene that plays a key role in cell cycle, transformation and proliferation and was selected because its knockdown is expected to cause similar response across cancer cell lines. The smaller overall distance between cell lines in Figure 4B indicates that the expected similar effect of MYC knockdown is better highlighted on a functional level between enriched biological processes rather than between transcriptomic signatures (Figure 4A). Furthermore, ligand-based approaches for vHTS are based on the hypothesis that similar chemical structures will bind to the same target and cause similar biological response. This hypothesis was validated computationally for structurally similar compounds in the CMap dataset, with biological response measured either at the gene or BP-level. For different thresholds, compound pairs with similar chemical structure were identified based on the Tanimoto distance between their ECFP4 fingerprints and the percentage of these pairs that also cause similar biological effect is presented in Table 1.[40] A comparison between structural and biological effect distances for all examined cell lines is presented in ESI 1.3. Across all structural distance thresholds the percentage of structurally similar compounds with similar biological effect is significantly higher when distance is calculated between signatures of enriched BPs. Thus, a distance metric calculated between enriched BPs, rather than between transcriptomic signatures, provides stronger evidence for support of the initial ligand-based hypothesis.

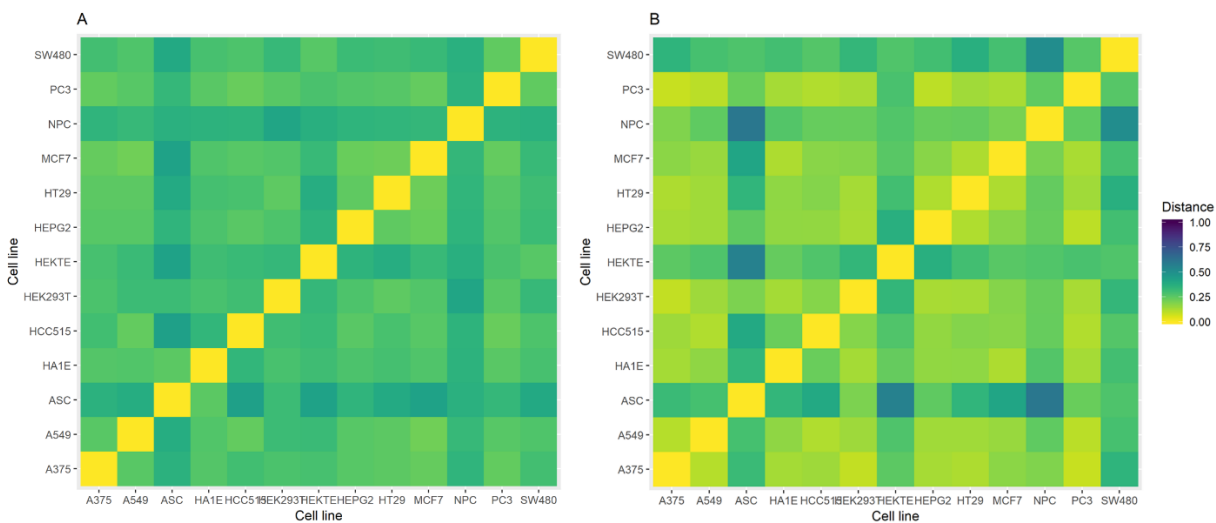

Figure 4 Heatmap of distance values between cell lines for the knockdown of the MYC gene. Across all cell lines, the transcriptomic signatures of MYC knockdown have Q1 score; (A) distances calculated between transcriptomic signatures; (B) distances calculated between enriched BPs.

Table 1 Percentage of structurally similar compounds with similar biological effect, either at the gene or BP-level, in the MCF7 cell line

| Structural distance threshold | Pairs of similar chemical structure | % of pairs affecting similar BPs* | % of pairs affecting similar genes** |
|---|---|---|---|
| 0.10 | 91 | 76.9 | 68.1 |
| 0.15 | 114 | 75.4 | 65.7 |
| 0.20 | 200 | 74.0 | 61.0 |
| 0.25 | 316 | 69.9 | 57.6 |
| 0.30 | 494 | 65.3 | 51.0 |

\* BP distance threshold 0.2
\*\* Gene distance threshold 0.19

To summarize, we showed that data quality greatly affects the distribution of the distance value, that distance between the same compounds' transcriptomic signatures differs across cellular models and that the similar biological effect of structurally similar compounds is better represented on a functional level. Based on these findings, the learning problem was formulated in such a way as to minimize the effect of the aforementioned factors. Thus, we chose to train cell line specific models for the A375, MCF7, PC3 and VCAP cell lines, using only compounds with Q1 GEx data, to predict the pairwise distance between enriched biological processes.

### 3.2 Performance evaluation

Model performance was evaluated on pairs of reference and test compounds. Test compounds were removed from the training sets and thus represent new chemical structures without available GEx data. Additionally, the effect of the structural similarity between input compounds on performance, along with the utility of the model's estimate for uncertainty, were investigated. Finally, we evaluated the performance of our approach on test chemical structures that are very different from the ones used in training.

**3.2.1 Test set performance.** In the MCF7 cell line, the performance of random initialization and deepSIBA augmented ensembles was compared to the performance of ReSimNet ensemble models and the results are presented in Table 2 (see ESI 4.2 for details). Furthermore, augmented ensembles were trained on data sets augmented with bootstrapped pairs between Q1 and Q2 signatures in an attempt to increase the chemical coverage of the training space (see ESI 3.2). The random initialization deepSIBA ensemble model achieved the lowest mean squared error (MSE) overall, while the augmented deepSIBA ensemble performed better in terms of precision and MSE in the 1% of test samples with the lowest predicted values. While ReSimNet ensembles had lower overall precision, they predicted that more compounds have similar biological effect. When examining the lowest 1% of predicted distances, ReSimNet's precision increases and becomes comparable to deepSIBA's. For the rest of the cell lines, random initialization deepSIBA ensembles were trained and tested on their respective data sets and their results are presented in Table 3. Additional 5-fold cross validation results for each cell line are presented in ESI 4.3. Random initialization ensembles were selected over augmented ensembles due to their reduced training time and reliability of Q1 signatures. Across all cell lines, deepSIBA ensembles produced consistently good results in their respective test sets.

Table 2 Model performance in the test set of the MCF7 cell line

| Model | MSE | MSE @1%* | Pearson's r. | Precision (%) | Precision @1% (%)* | No.Predicted Similars |
|---|---|---|---|---|---|---|
| DeepSIBA ensemble | **0.012** | 0.007 | 0.56 | 61.03 | 61.03 | 195 |
| ReSimNet ensemble | 0.015 | 0.029 | **0.59** | 26.93 | 51.20 | 13420 |
| DeepSIBA ensemble (augmented) | 0.014 | **0.005** | 0.49 | **74.63** | **74.63** | 205 |
| ReSimNet ensemble (augmented) | 0.014 | 0.015 | 0.47 | 39.47 | 53.40 | 3068 |

* Calculated at the 1% of test samples, with the lowest predicted values

Table 3 Cell line specific test set performance of deepSIBA

| Cell-line | MSE | MSE @1%* | Pearson's r | Precision (%) | No.Predicted Similars |
|---|---|---|---|---|---|
| A375 | 0.008 | 0.006 | 0.59 | 98.22 | 169 |
| VCAP | 0.033 | 0.026 | 0.41 | 71.63 | 141 |
| PC3 | 0.011 | 0.007 | 0.53 | 89.29 | 28 |
| MCF7 | 0.012 | 0.007 | 0.56 | 61.03 | 195 |

* Calculated at the 1% of test samples, with the lowest predicted values

**3.2.2 Performance as a function of the structural distance between input compounds.** As shown previously, similar chemical structures have similar signatures of enriched BPs. However, there are many cases of structurally dissimilar compounds that cause similar biological response. It is therefore important to evaluate the performance of our approach for test pairs of varying structural distance. On this front, each cell line specific test set was split into parts based on the distance between compounds' ECFP4 fingerprints and in each part MSE and precision were calculated (Figures 5A and 5B). The PC3, A375 and VCAP deepSIBA models maintain a high precision across all different structural distance ranges. The exception is the MCF7 model, for which precision slightly decreases for structural distance higher than 0.6. Regarding MSE, only the VCAP model exhibits a higher MSE as structural distance increases. As a whole, the models' performance seems unaffected by the distance between the ECFP4 fingerprints of the input pairs.

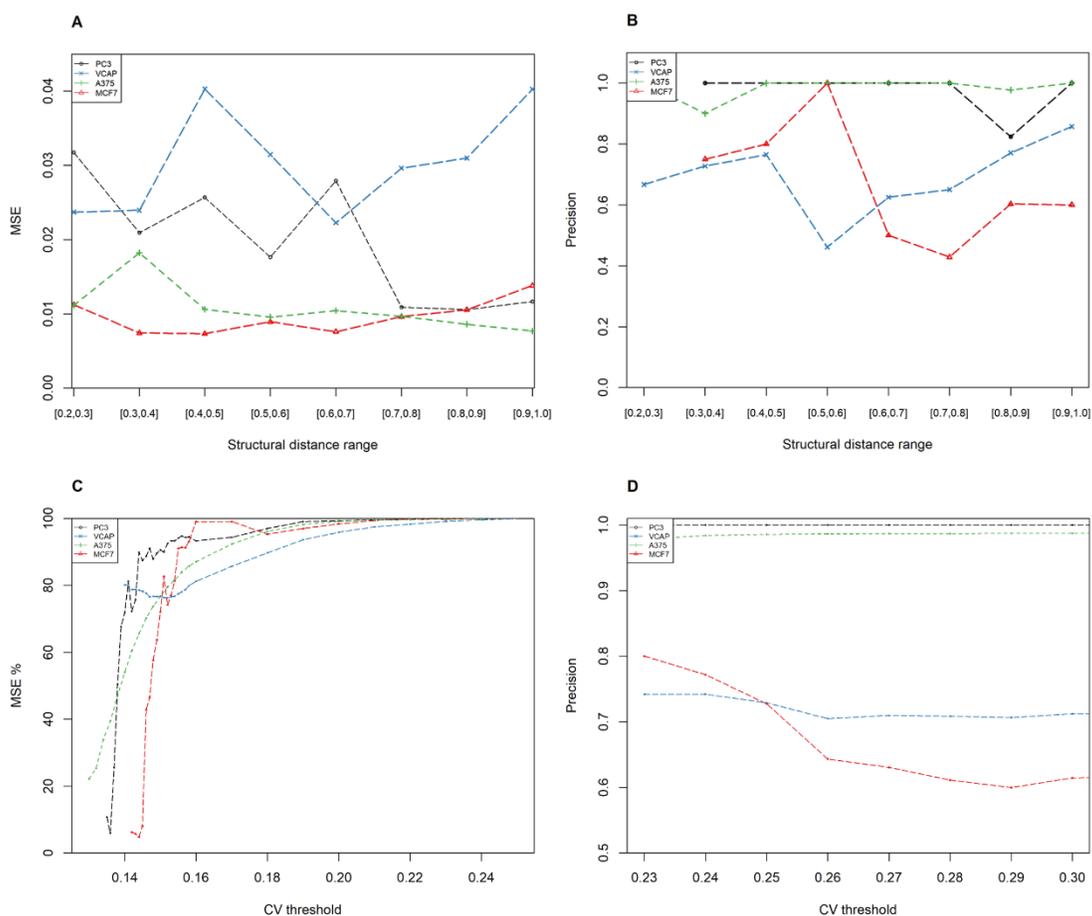

Figure 5 Performance as a function of structural distance and predictive uncertainty; (A) MSE for different ranges of structural distance between compound pairs; (B) Precision for different ranges of structural distance between compound pairs. (C) Percentage of total test MSE, calculated in samples with CV lower than an increasing threshold; (D) Precision calculated in test samples with CV lower than an increasing threshold.

**3.2.3 Performance as a function of predictive uncertainty.** It has been shown that quantifying predictive uncertainty can lead to more accurate results in virtual screening applications.[31] In this context, the relationship between the predictive uncertainty estimate and performance was investigated. In DeepSIBA we estimate predictive uncertainty as the coefficient of variation (CV) of the mixture of each model's Gaussian in the ensemble. MSE and precision were calculated for specific samples in the test set, which have CV lower than an increasing threshold and are presented in Figures 5C and 5D. As the CV threshold increases and more samples with higher CV are included in the evaluation, the MSE of the models increases as well and eventually becomes the MSE of the entire test set (Figure 5C). On the other hand, due to the low number of false positives, for all the models, precision seems unaffected by the CV threshold. Only the MCF7 model, which has the lowest overall precision, exhibits a higher precision for samples with lower CV. Overall, the results indicate that point predictions with lower uncertainty are closer to the true value, or that when the model is certain, it's usually not wrong.

**3.2.4 Generalization on different chemical structures.** End-to-end deep learning models for drug discovery have trouble generalizing on new compounds that are structurally very different from the ones used to train them. In order to evaluate the ability of our approach to generalize on different chemical structures, the performance of the A375 model was evaluated on 2 extra test sets and is presented in Table 4. These test sets were created by restricting the maximum allowed structural similarity between selected test compounds and all remaining training compounds (Figure 6A). As the minimum distance between test and training compounds increases, the performance of the model becomes worse. However, the performance decrease in terms of MSE and Pearson's r is smaller than the decrease in precision. In this case, the distance threshold to calculate precision was set to 0.22, because in the test set #3 there were no samples with predicted value lower than 0.2. Thus, even though the model's performance is comparable across test sets in terms of regression metrics, its ability to identify compounds with similar biological effect is hindered. In this case, it is important to estimate predictive uncertainty and evaluate its utility, by focusing on predictions with smaller CV (Figure 6B). In test set #3, which only contains compounds with maximum similarity to the training compounds less than 0.3, the model's precision is significantly higher for test predictions with low CV. More specifically, in test samples with CV lower than 0.16, the model's precision is upwards of 80%.

Table 4 Generalization performance on different chemical structures for A375

| Test set id | Range of max. similarity to training set | MSE | Pearson's r | Precision (%) | No.Predicted Similars |
|---|---|---|---|---|---|
| #1 | [0-0.85] | 0.0083 | 0.59 | 97.26 | 876 |
| #2 | [0.35-0.65] | 0.0092 | 0.52 | 76.48 | 330 |
| #3 | [0-0.3] | 0.0107 | 0.44 | 50.37 | 135 |

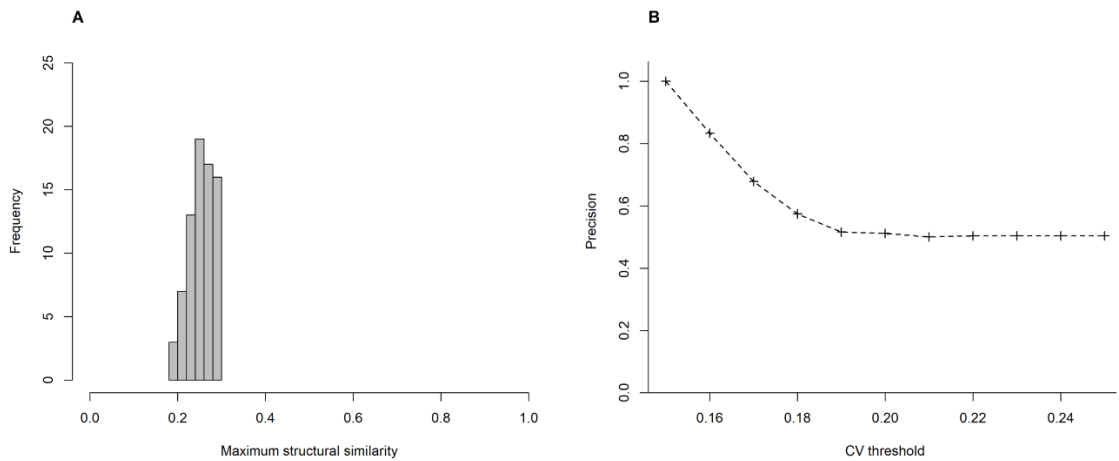

Figure 6 Precision and uncertainty estimation for test set #3. (A) Histogram of maximum structural similarity between test and training compounds for test set #3; structural similarity is calculated between compounds' ECFP4 fingerprints. (B) Precision calculated in test samples with CV lower than an increasing threshold.

Across all examined cell lines, deepSIBA was able to identify chemical structures that affect similar BPs with high precision and low MSE, outperforming the architecture that utilizes compounds' ECFP4 fingerprints as input. Furthermore, the models were able to identify structurally dissimilar compounds with similar biological effect that would have been omitted by traditional fingerprint similarity algorithms. This shows that the employed GCNN architecture was able to learn new meaningful representations of chemical structures related to their biological effect. However, there were some cases of compounds affecting similar BPs that were missed by the model. These cases, in combination with the decrease in performance as the minimum distance between test and training compounds increases highlight key limitations in our approach. On this front, limited coverage of the chemical space by compounds with available GEx data is a major issue that limits our ability to model in its entirety the complex function that translates changes in chemical structure to BP alterations. Even though each training set for each cell line contains on average around 320K samples, these are comprised from the pairing of around 800 compounds. The limitations that arise from this low coverage of the chemical space can't be solved by changes in deep learning architecture and require more training compounds and/or extra input information. One efficient workaround that we utilized in our approach is to quantify predictive uncertainty using deep ensembles. We showed that the model's performance, even when tested on compounds that are structurally different from the ones used in training, is higher for samples with lower uncertainty. Thus, the model's estimate of uncertainty can be used to provide more reliable and accurate results. For example, if an application imposes a constraint on the maximum allowed error, the appropriate uncertainty threshold can be identified and only point predictions with uncertainty lower than this threshold be taken into account.

## 3.3 Signaling pathway inference for target structure

The predictions of deepSIBA can be used to infer a signaling pathway signature, in terms of the most upregulated and downregulated pathways, for a target chemical structure without available GEx data. The inference is performed following a KNN-like approach, in which reference compounds with the smallest distance to the target, as predicted by the model, are selected as its neighbors and their pathway signatures are retrieved. Then, pathways that frequently belong in the 10 most upregulated or downregulated pathways of the neighbors are inferred as the target's signature. The performance of the approach was evaluated on the test compounds of the MCF7 model and then, as a use case, it was tasked to infer the signaling pathways affected by the FDA approved anticancer drugs, for which no GEx data are available in our dataset.

**3.3.1 Performance evaluation in the test set of MCF7**. For the test set of the MCF7 cell line, the average performance of the inference approach is presented in Table 5. On average 5 pathways per test compound were inferred to belong in its 10 most downregulated pathways with a precision of 73.3%. Regarding upregulation, an average of 2.5 pathways per compound with a precision of 69.7% were inferred. We have to note that the statistical significance of the inferred pathways is ensured by comparing the neighbor selection process using the trained model to a random selection.

Table 5 Average pathway inference results for the test compounds of MCF7

|               | Number of inferred pathways | Precision (%) |
|---------------|:---------------------------:|:-------------:|
| Downregulated | 5                           | 73.3          |
| Upregulated   | 2.5                         | 69.7          |

**3.3.2 Use case: signaling pathway inference of FDA approved anticancer drugs**. Out of the 59 FDA-approved cytotoxic drugs presented by Sun *et al.*, 18 were present or had a structural analogue in the MCF7 training set (Tanimoto ECFP4 similarity > 0.85).[41] In order to simulate a realistic application for the signaling pathway inference, these 18 drugs were excluded from the target set. From the remaining 39 drugs, only 3 had more than 5 neighbors in the training set, as predicted by the model and their inferred pathways are presented in Table 6. Fludarabine and Clofarabine are direct nucleic acid synthesis inhibitors, while Pralatrexate is an indirect inhibitor of nucleotide synthesis through inhibition of the folate cycle.[42] In our use case, the inferred downregulated signaling pathways include cell cycle, purine and pyrimidine metabolism, RNA transport and spliceosome, which are closely related to the drugs' mechanism of action. Furthermore, because of the MCF7 cell line, pathways such as oocyte meiosis and progesterone-mediated oocyte maturation, that have been associated with the pathogenesis of breast cancer, were inferred as downregulated.[43] Regarding upregulation, pathways such as NF-kappa B signaling, natural killer cell mediated cytotoxicity, leukocyte transendothelial migration and TNF signaling, that are closely related to inflammation and apoptosis, were inferred.

Table 6 Pathway inference results for FDA approved anticancer drugs

| Drug | Mechanism of Action | Inferred Downregulated KEGG Signaling Pathways | Inferred Upregulated KEGG Signaling Pathways |
|---|---|---|---|
| Fludarabine | Nucleic Acid Synthesis Inhibitor | **Purine metabolism, Pyrimidine metabolism**, **RNA transport**, **Spliceosome**, **Cell cycle**, **Oocyte meiosis**, **Progesterone-mediated oocyte maturation,** MicroRNAs in cancer | Leukocyte transendothelial migration, Oxytocin signaling pathway, Alzheimer's disease, Pertussis, Rheumatoid arthritis |
| Clofarabine | Nucleic Acid Synthesis Inhibitor | **RNA transport**, **Spliceosome**, **Cell cycle**, Ubiquitin mediated proteolysis, **Progesterone-mediated oocyte maturation**, MicroRNAs in cancer | **Natural killer cell mediated cytotoxicity**, Leukocyte transendothelial migration, Oxytocin signaling pathway, Pertussis, Rheumatoid arthritis |
| Pralatrexate | Inhibits dihydrofolate reductase (DHFR) and thymidylate synthase | **Purine metabolism, Pyrimidine metabolism**, Metabolic pathways, **RNA transport, Spliceosome** | **NF-kappa B signaling pathway, Natural killer cell mediated cytotoxicity, TNF signaling pathway**, Leukocyte transendothelial migration |

We demonstrated that by utilizing the training compounds as reference, the signaling pathway inference approach can provide an early estimate regarding the signaling pathways affected by an unknown compound. Due to the nature of the inference method, limiting factors can arise from the lack of diversity in affected BPs by the training compounds. This lack of diversity can influence the signaling pathway inference for an unknown target structure, when its true biological effect is vastly different from the effect of all training compounds. In such cases, it is important to avoid the inference of incorrect pathways as being downregulated or upregulated and infer statistically significant pathways for target compounds that have at least k neighbors in the reference set. This was apparent in the presented use case, where out of the 39 compounds a pathway signature was inferred only for 3, but for those the inferred pathways were directly related to their mechanism of action.

# 4 Conclusion

In this paper, we developed a deep learning framework to match the chemical structure of compound perturbations to their biological effect on specific cellular models. We showed, that the careful formulation of the learning problem and the flexibility of the Siamese GCNN architecture enabled our models to achieve high performance across all test scenarios. Additionally, we highlighted the utility of the uncertainty estimate, provided by deep ensembles, in test cases where the unknown chemical structures are very different from the structures used to train the models. Finally, we presented a novel inference pipeline, which can infer a signaling pathway signature for a target compound without available GEx data, using only its molecular structure. The high performance of our methods paves the way for further investigation in order to expand their coverage and utility.

Possible efforts for further investigation can be concentrated on the input representation, the biological response distance and the model's uncertainty estimate. Regarding the input, one interesting idea is to include binding information in order to capture the potential protein target on the input molecules. This extra information can be passed to the model either in the form of the latent space embeddings from a trained binding affinity prediction model or in the form of predictions against a panel of protein kinases.[44] Regarding the biological distance between compound perturbations, this can be augmented by calculating the compound's effect on different levels of biological hierarchy, i.e. GEx, signaling pathways, transcription factors and signaling networks.[45, 46] Afterwards, these distances could be combined or separate models could be trained in order to better capture the similar effect of compounds. Additionally, instead of using a distance metric between all affected BPs, specific biological processes could be selected and application specific models could be developed to identify compounds that affect these biological processes. Finally, regarding the model's uncertainty estimate, an interesting avenue for investigation is to take into account the transcriptomic signatures of replicates from the CMap dataset and calculate distributions of pairwise distances between compounds. Then, models could be trained on these distributions to better capture the variation of the experimental ground truth.

The framework that we developed borrows aspects from both CADD methods and systems pharmacology-based approaches in order to incorporate the structural and systematic effects of small molecule perturbations, which are closely related to their efficacy and toxicity profiles. We believe that our methods have the potential to augment vHTS approaches for hit identification, either by reasoning on a massive scale about the biological effect of compounds, without available GEx data, or by suggesting new chemical structures with desired biological effect.

## Data and code availability

All analyzed data that were used to train our models and produce all tables and figures are available at https://github.com/BioSysLab/deepSIBA. Furthermore, the R source code to analyze the CMap dataset and create the training, validation and test sets is available at https://github.com/BioSysLab/deepSIBA. Finally, the Keras/TensorFlow implementation of our deep learning models, alongside trained ensemble models for each cell line are available at https://github.com/BioSysLab/deepSIBA.

## Conflicts of interest

There are no conflicts to declare.

## Acknowledgements

The authors would like to thank T. Sakellaropoulos and M. Neidlin for their constructive feedback on the manuscript. This research has been co-financed by the European Union and Greek national funds through the Operational Program Competitiveness, Entrepreneurship and Innovation, under the call RESEARCH – CREATE – INNOVATE (project code:T1EDK-02829).

# References

1. G. Sliwoski, S. Kothiwale, J. Meiler and E. W. Lowe, *Pharmacological reviews*, 2014, **66**, 334-395.
2. J. Hert, P. Willett, D. J. Wilton, P. Acklin, K. Azzaoui, E. Jacoby and A. Schuffenhauer, *Journal of chemical information and modeling*, 2006, **46**, 462-470.
3. F. Sirci, F. Napolitano, S. Pisonero-Vaquero, D. Carrella, D. L. Medina and D. di Bernardo, *NPJ systems biology and applications*, 2017, **3**, 1-12.
4. J. P. Bai and D. R. Abernethy, *Annual review of pharmacology and toxicology*, 2013, **53**, 451-473.
5. C. Fotis, A. Antoranz, D. Hatziavramidis, T. Sakellaropoulos and L. G. Alexopoulos, *Drug discovery today*, 2018, **23**, 626-635.
6. B. Verbist, G. Klambauer, L. Vervoort, W. Talloen, Z. Shkedy, O. Thas, A. Bender, H. W. Göhlmann, S. Hochreiter and Q. Consortium, *Drug discovery today*, 2015, **20**, 505-513.
7. F. Iorio, R. Bosotti, E. Scacheri, V. Belcastro, P. Mithbaokar, R. Ferriero, L. Murino, R. Tagliaferri, N. Brunetti-Pierri and A. Isacchi, *Proceedings of the National Academy of Sciences*, 2010, **107**, 14621-14626.
8. J. Lamb, E. D. Crawford, D. Peck, J. W. Modell, I. C. Blat, M. J. Wrobel, J. Lerner, J.-P. Brunet, A. Subramanian and K. N. Ross, *science*, 2006, **313**, 1929-1935.
9. A. Subramanian, R. Narayan, S. M. Corsello, D. D. Peck, T. E. Natoli, X. Lu, J. Gould, J. F. Davis, A. A. Tubelli and J. K. Asiedu, *Cell*, 2017, **171**, 1437-1452. e1417.
10. S. Pushpakom, F. Iorio, P. A. Eyers, K. J. Escott, S. Hopper, A. Wells, A. Doig, T. Guilliams, J. Latimer and C. McNamee, *Nature reviews Drug discovery*, 2019, **18**, 41-58.
11. J. Vamathevan, D. Clark, P. Czodrowski, I. Dunham, E. Ferran, G. Lee, B. Li, A. Madabhushi, P. Shah and M. Spitzer, *Nature Reviews Drug Discovery*, 2019, **18**, 463-477.
12. H. Chen, O. Engkvist, Y. Wang, M. Olivecrona and T. Blaschke, *Drug discovery today*, 2018, **23**, 1241-1250.
13. H. Öztürk, A. Özgür and E. Ozkirimli, *Bioinformatics*, 2018, **34**, i821-i829.
14. W. Jin, R. Barzilay and T. Jaakkola, *arXiv preprint arXiv:1802.04364*, 2018.
15. DeepChem: Deep-learning models for Drug Discovery and Quantum Chemistry, (accessed 2020-03-20).
16. Z. Wu, B. Ramsundar, E. N. Feinberg, J. Gomes, C. Geniesse, A. S. Pappu, K. Leswing and V. Pande, *Chemical science*, 2018, **9**, 513-530.
17. Y. LeCun, Y. Bengio and G. Hinton, *nature*, 2015, **521**, 436-444.
18. D. K. Duvenaud, D. Maclaurin, J. Iparraguirre, R. Bombarell, T. Hirzel, A. Aspuru-Guzik and R. P. Adams, *Journal*, 2015, 2224-2232.
19. W. Torng and R. B. Altman, *Journal of Chemical Information and Modeling*, 2019, **59**, 4131-4149.
20. S. Kearnes, K. McCloskey, M. Berndl, V. Pande and P. Riley, *Journal of computer-aided molecular design*, 2016, **30**, 595-608.
21. I. Wallach and A. Heifets, *Journal of chemical information and modeling*, 2018, **58**, 916-932.
22. F. Schroff, D. Kalenichenko and J. Philbin, 2015.
23. L. Bertinetto, J. Valmadre, J. F. Henriques, A. Vedaldi and P. H. Torr, 2016.
24. O. Vinyals, C. Blundell, T. Lillicrap and D. Wierstra, 2016.
25. Y. Bai, H. Ding, Y. Sun and W. Wang, *arXiv preprint arXiv:1810.10866*, 2018.
26. H. Altae-Tran, B. Ramsundar, A. S. Pappu and V. Pande, *ACS central science*, 2017, **3**, 283-293.
27. Y. Gal and Z. Ghahramani, 2016.
28. B. Lakshminarayanan, A. Pritzel and C. Blundell, 2017.
29. S. Jain, G. Liu, J. Mueller and D. Gifford, *arXiv preprint arXiv:1906.07380*, 2019.

30. A. Kendall and Y. Gal, 2017.
31. S. Ryu, Y. Kwon and W. Y. Kim, *Chemical Science*, 2019, **10**, 8438-8446.
32. M. Jeon, D. Park, J. Lee, H. Jeon, M. Ko, S. Kim, Y. Choi, A.-C. Tan and J. Kang, *Bioinformatics*, 2019, **35**, 5249-5256.
33. GEO GSE92742, https://www.ncbi.nlm.nih.gov/geo/query/acc.cgi?acc=GSE92742, (accessed 2020-03-20).
34. CLUE platform, https://clue.io/, (accessed 2020-03-20).
35. A. Alexa and J. Rahnenführer, *Bioconductor Improv*, 2009, **27**.
36. A. Sergushichev, *BioRxiv*, 2016, 060012.
37. F. Li, Y. Cao, L. Han, X. Cui, D. Xie, S. Wang and X. Bo, *Omics: a journal of integrative biology*, 2013, **17**, 116-118.
38. M. Kanehisa, 2002.
39. M. Iwata, R. Sawada, H. Iwata, M. Kotera and Y. Yamanishi, *Scientific reports*, 2017, **7**, 40164.
40. D. Rogers and M. Hahn, *Journal of chemical information and modeling*, 2010, **50**, 742-754.
41. J. Sun, Q. Wei, Y. Zhou, J. Wang, Q. Liu and H. Xu, *BMC systems biology*, 2017, **11**, 87.
42. O. Shuvalov, A. Petukhov, A. Daks, O. Fedorova, E. Vasileva and N. A. Barlev, *Oncotarget*, 2017, **8**, 23955.
43. D. Wu, B. Han, L. Guo and Z. Fan, *Journal of Obstetrics and Gynaecology*, 2016, **36**, 615-621.
44. A. Cichonska, B. Ravikumar, R. J. Allaway, S. Park, F. Wan, O. Isayev, S. Li, M. Mason, A. Lamb and Z. Tanoli.
45. L. Garcia-Alonso, F. Iorio, A. Matchan, N. Fonseca, P. Jaaks, G. Peat, M. Pignatelli, F. Falcone, C. H. Benes and I. Dunham, *Cancer research*, 2018, **78**, 769-780.
46. A. Liu, P. Trairatphisan, E. Gjerga, A. Didangelos, J. Barratt and J. Saez-Rodriguez, *NPJ systems biology and applications*, 2019, **5**, 1-10.

# Electronic Supplementary Information

## S1 Data preprocessing and quality control

**S1.1 Dataset and quality overview.** The filtered CMap dataset contains 112994 transcriptomic signatures from 20254 compounds tested across 70 cell lines. During the filtering process, for each compound per cell line, its signature with the highest quality across different dosages and time points was selected. The assigned quality score based on TAS, number of replicates and whether the signature is considered an exemplar is presented in Table S1. The distribution of signature quality across cell lines is presented in Figure S1. Deep learning models were developed for the MCF7, PC3, VCAP and A375 cell lines, which have the highest number of compounds with Q1 signatures.

**Table S1** Signature quality score

| Quality score | TAS | Number of replicates | Exemplar |
|---|---|---|---|
| Q1 | > 0.4 | > 2 | True |
| Q2 | 0.2 – 0.4 | > 2 | True |
| Q3 | 0.2 – 0.4 | ≤ 2 | True |
| Q4 | 0.2 – 0.4 | > 2 | True |
| Q5 | 0.2 – 0.4 | ≤ 2 | True |
| Q6 | < 0.1 | > 2 | True |
| Q7 | < 0.1 | ≤ 2 | True |
| Q8 | < 0.1 | < 2 | False |

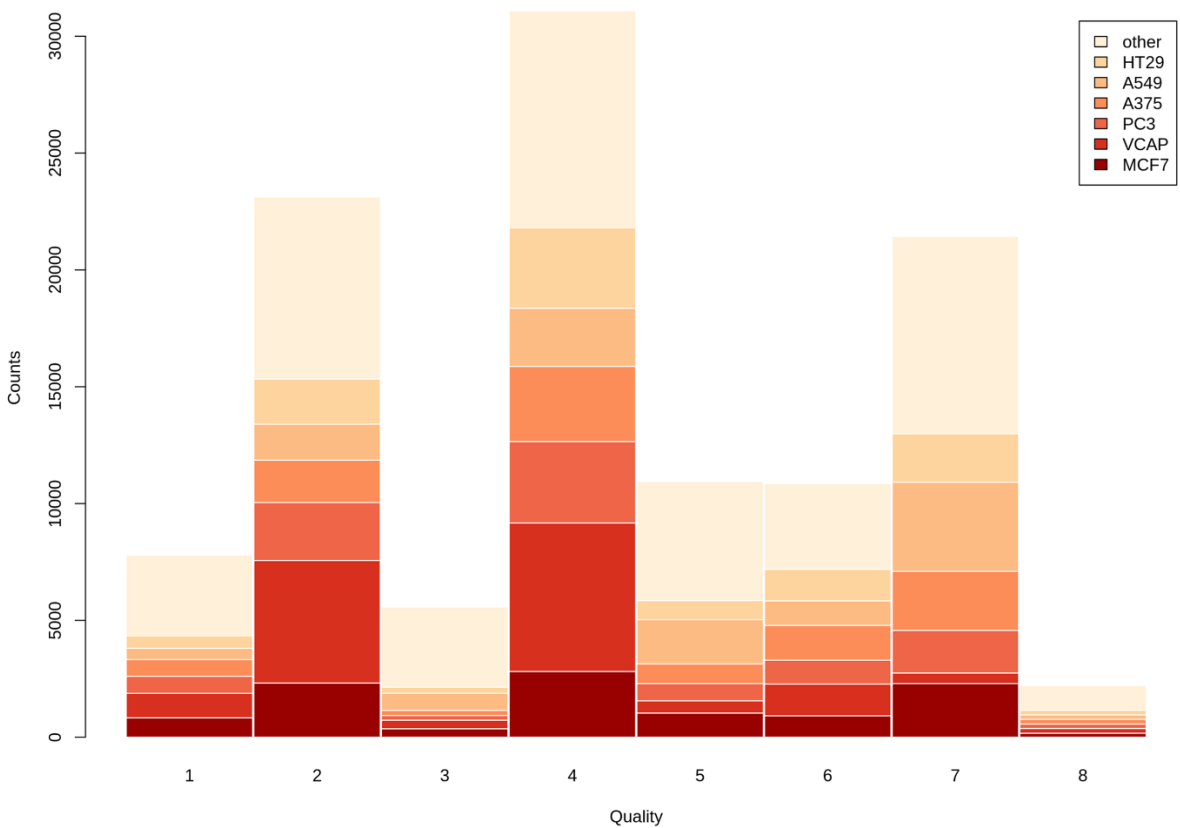

**Figure S1** Distribution of signature quality scores across cell lines. The "other" cell line category is formed by grouping together 63 cell lines with smaller number of available transcriptomic signatures.

**S1.2 GO term enrichment and distance calculation.** For the MCF7, A375, VCAP and PC3 cell lines, the average number of significantly enriched GO terms in quality 1 signatures is presented in Table S2. Enrichment p-values were calculated with GSEA and adjusted using the Benjamini-Hochberg procedure. GO terms with an adjusted p-value less than 0.05 were considered significantly enriched. Based on Table S2 the number of top and bottom GO terms to consider during the ensemble distance calculation was selected (10, 20, 30, 40 and 50 GO terms). The ensemble distance approach outputs 5 distance scores for each signature pair, one for each of the numbers of top and bottom GO terms considered. The histogram of standard deviations of the calculated distances for each cell line is presented in Figure S2. The effect of the number of GO terms to consider during distance calculation is small, but not negligible. Furthermore, the relationship between pairwise distances between compounds at the GO term-level and at the gene-level was examined (Figure S3). Although distances are significantly correlated, the similar biological effect of chemical structures is better represented on a functional level between enriched GO terms. Finally, the ensemble distance approach of the GO term feature vectors was validated computationally. For the MCF7 cell line, where enough quality 1 duplicate signatures are available (n = 20), their distance distribution was compared to a randomly selected subset of pairwise distances between different compound perturbations (Figure S4). It is clear that the proposed ensemble distance metric can easily separate duplicate compound perturbation pairs from random pairs.

**Table S2** Average number of significantly enriched GO terms following compound treatment

| Cell line | Average number of significant GO terms (p.adj < 0.05) |
|---|---|
| MCF7 | 20.1 |
| A375 | 29.8 |
| VCAP | 11.2 |
| PC3 | 30.0 |

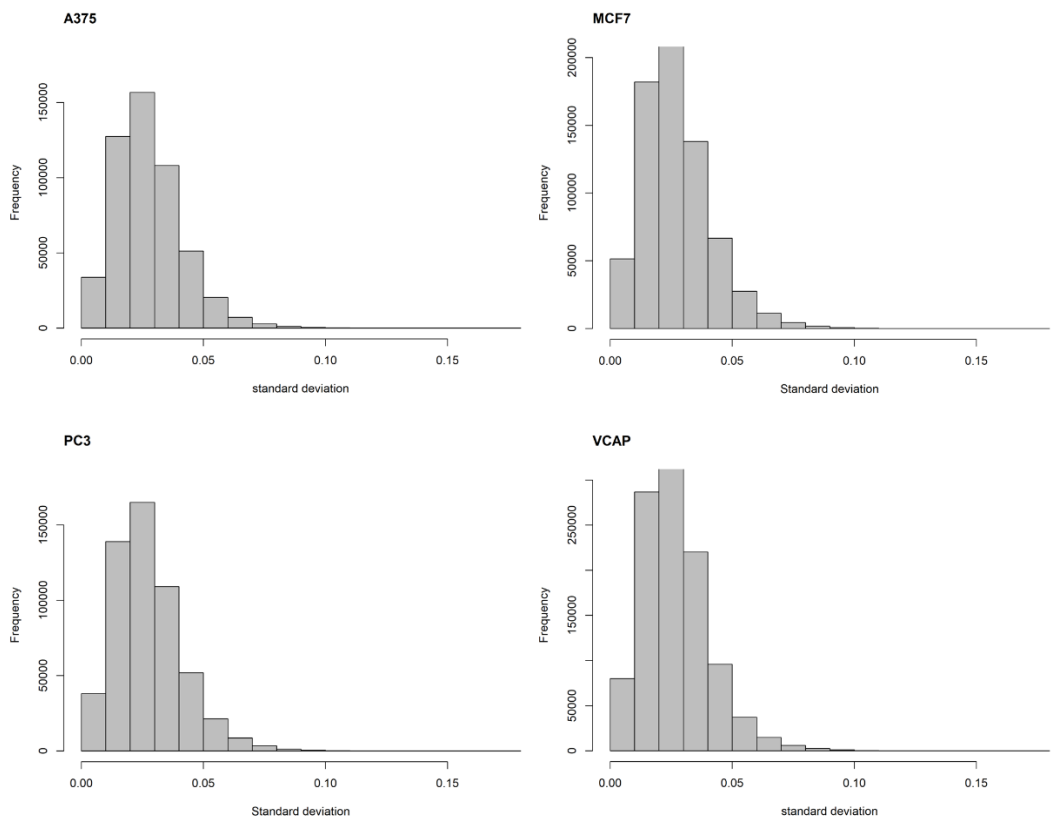

**Figure S2** Histograms of standard deviations of distances calculated between enriched GO terms for 5 different numbers of top and bottom GO terms (10, 20, 30, 40 and 50) for each cell line. Distances were calculated between compounds with Q1 signatures only.

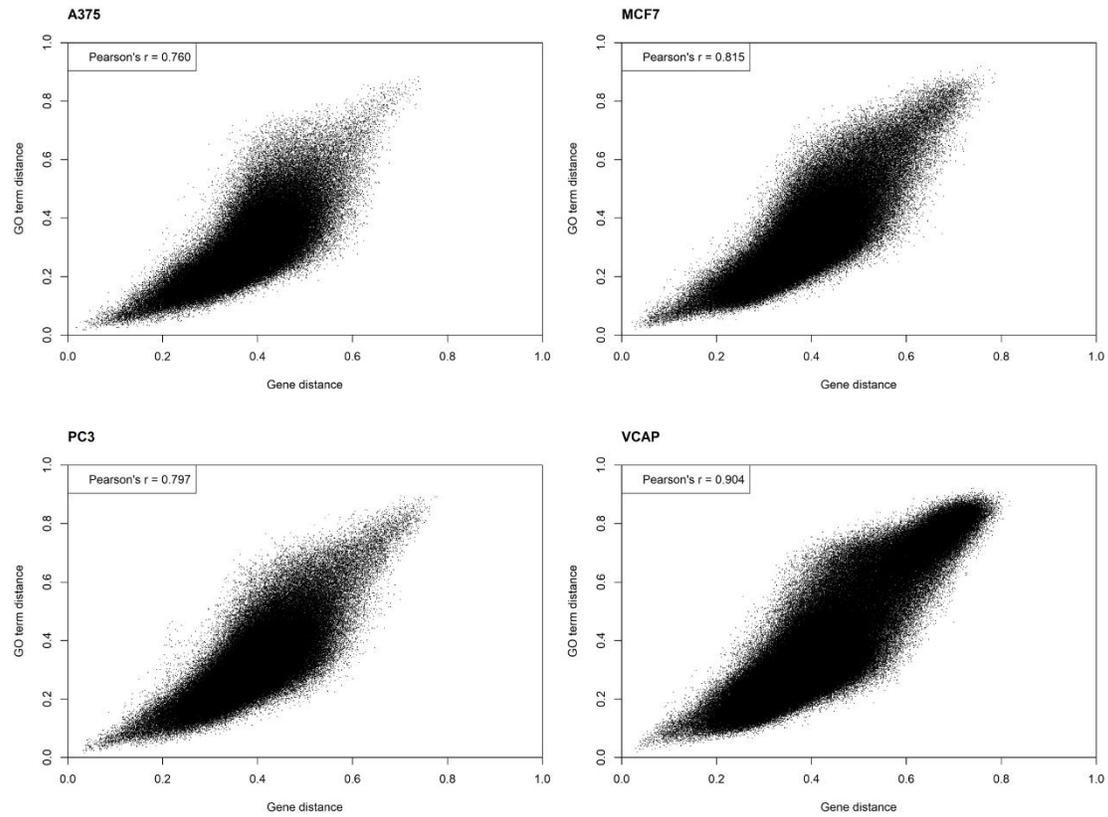

**Figure S3** Scatter plot of pairwise distances between compounds calculated at the gene and GO term-level for each cell line. Distances were calculated between compounds with Q1 signatures only.

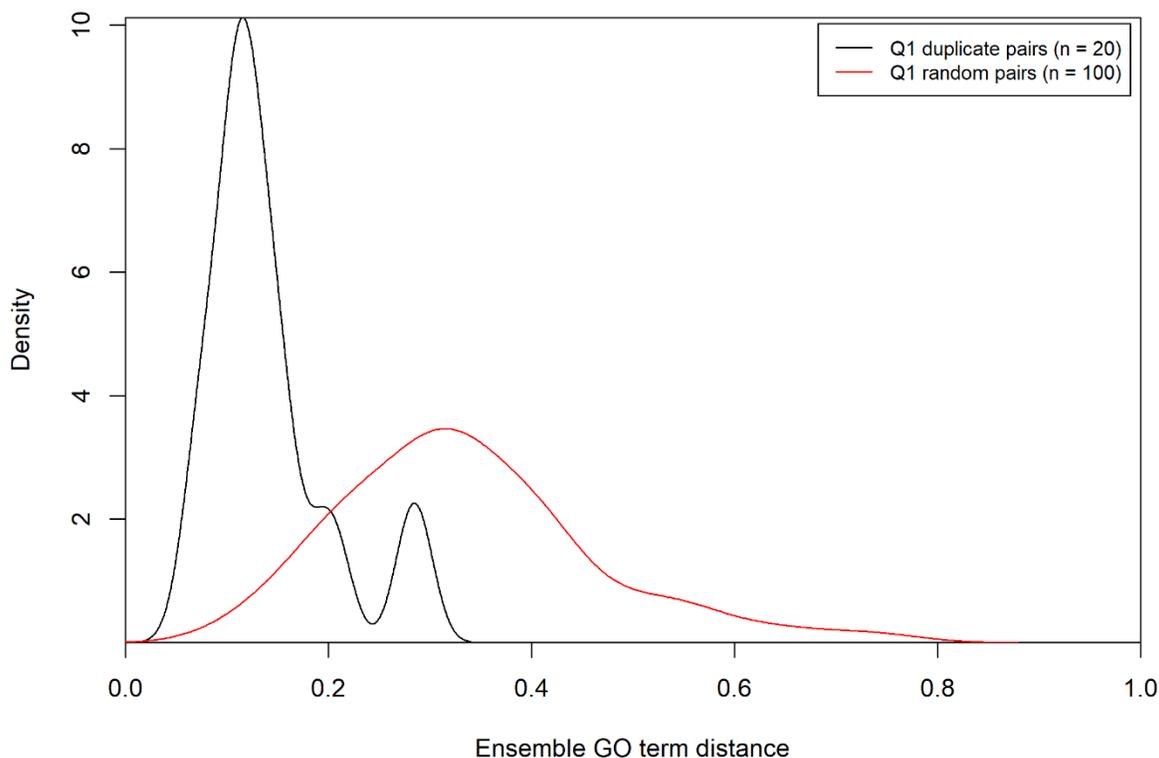

**Figure S4** Distribution of distances calculated with the ensemble GSEA score approach between compounds' affected BPs for the MCF7 cell line. The black line represents the distribution of pairwise distances between duplicate signatures, while the red line represents the distances between signatures of random compound pairs. The separation between the two distributions indicates that the ensemble distance function can distinguish compounds that affect similar BPs (duplicates) from random compound pairs.

**S1.3 Comparing structural and biological effect distance.** For each compound pair, Morgan circular fingerprints with radius 2 were generated using RDkit and their pairwise Tanimoto coefficient (Tc) was computed.[1] During fingerprint generation, the default atom invariants were used, making them similar to the widely used ECFP4.[2] Finally, pairwise compound distances were calculated, as $1 - Tc$. The relationship between pairwise compound structural distances and their distance in terms of affected biological processes (GO terms) in each cell line, was examined (Figure S5). We report similar results to Sirci *et al*.[3] and their analysis of transcriptomic and structural distances in the original CMAP dataset.[4] Indeed, compounds with similar ECFP4 fingerprints, tend to affect similar biological processes (lower left quadrant of Figure S5). However, there are many structurally dissimilar compounds that have similar biological footprint (upper left quadrant of Figure S5). Finally, the majority of compounds are structurally dissimilar and affect different biological processes (upper right quadrant of Figure S5).

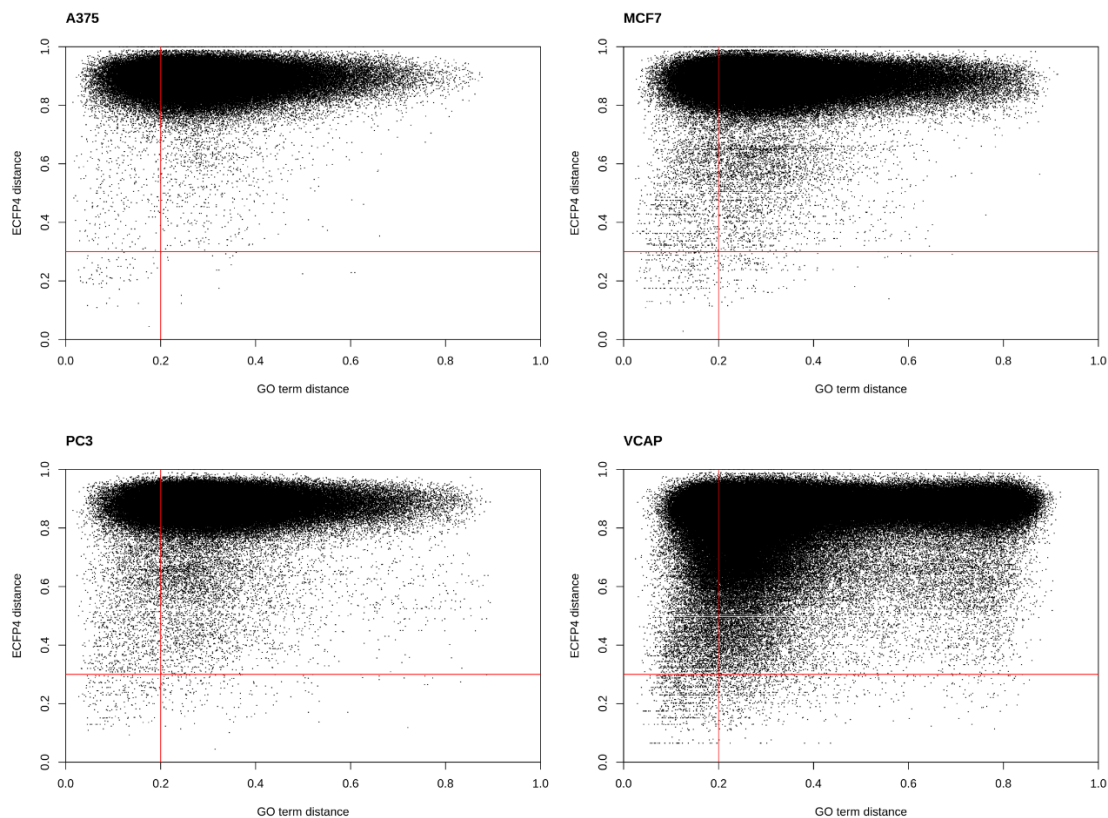

**Figure S5** Scatterplot of pairwise distances between compounds' ECFP4 fingerprints and between compounds' enriched BPs. The red lines represent reasonable thresholds to consider compounds similar in structure and similar in effect (0.3 for ECFP4 and 0.2 for BPs). Even though there is no correlation between compounds' structural and biological effect distances, the majority of structurally similar compounds tend to affect similar BPs (lower left quadrant of the plot).

## S2 Deep learning model

**S2.1 Input representation.** Molecular graphs are presented to the model using the Atom array, the Edge array and the Bond array. The Atom array has as many rows as the max number of atoms across all compounds and each column represents an atom feature. In total, 62 atom features are utilized. The atom features consist of the concatenated vectors of 4 one hot encoded features and 1 binary feature, which describe:

- The symbol of the atom (one-hot).
- The degree of the atom (one-hot).
- The number of attached hydrogen atoms (one-hot).
- The valence of the atom (one-hot).
- If the atom is aromatic (binary).

The Edge array describes the connectivity of the graph representing the molecule. The Edge array consists of as many rows as the max number of atoms. Each row contains the atom's neighbors. The Bond array is 3-dimensional and contains the features of each bond. Each row represents an atom, while each column represents a neighbor, up to 5 for each atom. A bond is described by 6 binary features contained in the Bond array, which describe whether the bond is:

- Single
- Double
- Triple
- Aromatic
- Conjugated
- In a Ring

The Atom, Bond and Edge arrays were created using RDKit in python.

**S2.2 Graph convolutions.** Graph convolutions were implemented in Keras as described by Duvenaud *et al.*[5]. A graph convolutional layer aggregates information from the neighboring nodes of a node/atom in the molecular graph. For every atom, its bond features are summed and concatenated with its atom feature vector. The resulting feature vector of each atom is summed with the feature vectors of its neighbors, using the connectivity information of the Edge array, creating in this way a new feature vector for every atom with aggregated information from the atom's neighborhood. Then, every feature vector passes through a fully connected layer, based on the atom's degree, and a non-linear activation function. Typically, following a graph convolution layer, a function, such as sum, is used to aggregate node embeddings into whole graph embeddings. In our implementation we omitted the use of an aggregation function and instead utilized 1D convolutions to gather information across neighborhoods and produce a graph feature map.

**Graph Convolutional Layer Pseudocode:**

1: **Input:** Atom array $X_A$, Bond array $X_B$, Edge array $D$

2: **for** each atom $a_i$ in a molecule

3: $\quad SX_{B_i} = \sum X_{B_i}$

4: $\quad X'_{A_i} = concatenate(X_{A_i}, SX_{B_i})$

5: $\quad$ **for** each neighbor j from N neighbors

6: $\quad\quad SX_{B_j} = \sum X_{B_j}$

7: $\quad\quad X'_{A_j} = concatenate(X_{A_j}, SX_{B_j})$

8: $\quad X''_{A_i} = X'_{A_i} + \sum_{j=1}^{N} X'_{A_j}$

9: $\quad X^{new}_{A_i} = relu(W_{degree} * X''_{A_i} + b_{degree})$ #is the new concatenated atom and bond matrix

**S2.3 Model hyperparameters.** The hyperparameters used to train the models are presented in Table S3. In our approach we utilized widely accepted hyperparameter values without performing hyperparameter optimization.

**Table S3** Model hyperparameters

| | |
|---|---|
| Optimizer | Adam |
| Learning Rate | 0.001 |
| Epochs | 20 |
| Batch size | 128 |
| Regularization | Dropout (rate = 0.3) |
| Batch Normalization Momentum | 0.6 |
| Weight Initializer | Glorot Normal |
| Activation Function | ReLu |

**S2.4 Gaussian mixture.** By using a Gaussian regression layer, each model outputs a mean and variance of the biological effect distance between pairs of molecular graphs. The ensemble's output is also a Gaussian, with mean and variance calculated from the uniformly weighted mixture of each model. The mean and variance of the mixture are defined as

$$m_u = \frac{1}{N}\sum_{i=1}^{N} \bar{y}_i$$

$$sigma = \sqrt{\frac{1}{N}\sum_{i=1}^{N}(sd_i + \bar{y}_i^2) - m_u^2}$$

where,

- $\bar{y}_i$ is the output mean value of the biological effect distance of each model.
- $N$ is the number of models.
- $m_u$ is the final mean value of the uniformly weighted mixture.
- $sigma$ is the standard deviation of the uniformly weighted mixture.
- $sd_i$ is the output variance of the biological effect distance of each model.

Finally, the coefficient of variation of the Gaussian mixture is used as the model's estimate of predictive uncertainty and is defined as

$$CV = \frac{sigma}{m_u}.$$

# S3 Dataset splitting and augmentation

**S3.1 Dataset splitting.** An overview of the training and test sets for each cell line is presented in Table S4 and S5, while the distribution of the target variable is presented in Figure S6. For the proposed learning task, random splitting of compound pairs between training and test has no benefit, since if a compound is present on the training set its affected BPs are known and distances to other compounds can be calculated instead of predicted.

**Table S4** Cell line specific training sets

| Cell line | Number of Compounds | Number of Pairs |
|---|---|---|
| MCF7 | 713 | 253828 |
| PC3 | 608 | 184528 |
| A375 | 592 | 174936 |
| VCAP | 934 | 435711 |

**Table S5** Cell line specific test sets

| Cell line | Number of Compounds | Number of Pairs |
|---|---|---|
| MCF7 | 70 | 49910 |
| PC3 | 74 | 44992 |
| A375 | 77 | 45584 |
| VCAP | 63 | 58842 |

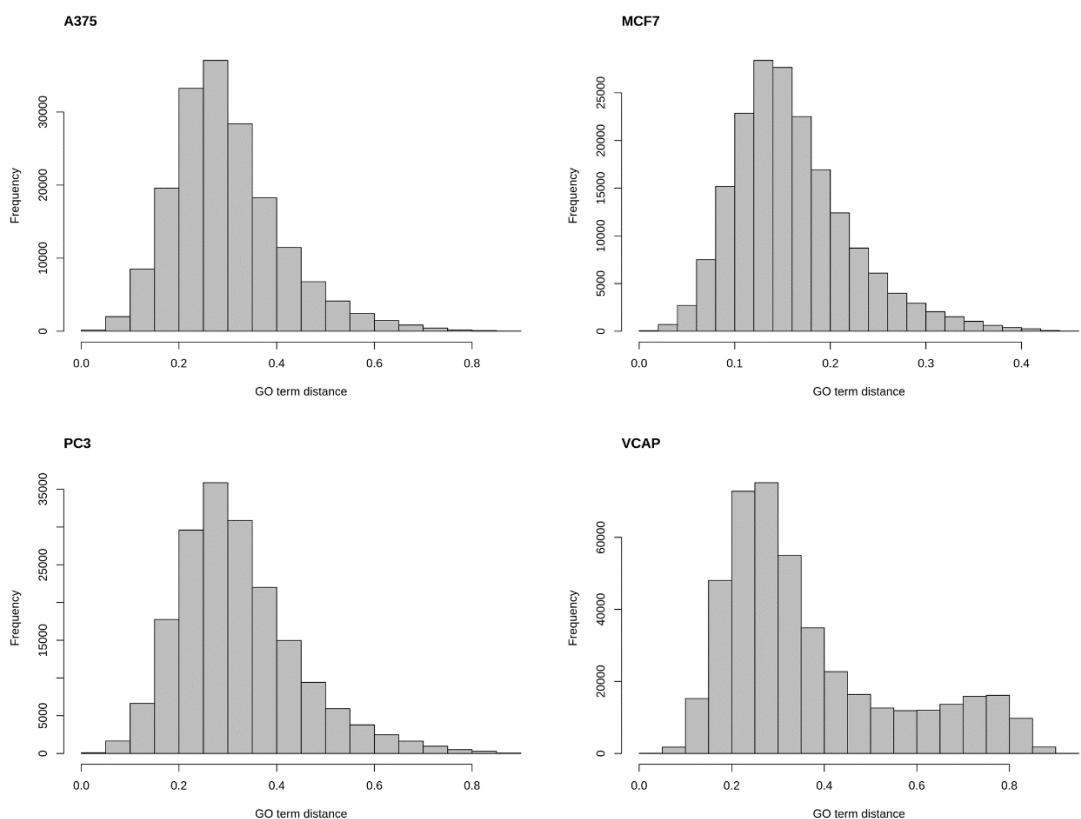

**Figure S6** Histogram of the model's target variable for each cell line.

**S3.2 Data Augmentation.** When training augmented ensemble models, the original training set of each model, consisting of Q1 signature pairs, was augmented with randomly sampled pairs between Q1 and Q2 signatures. This technique was utilized in order to increase the number of compounds and the diversity of chemical structures available during training. Although this approach resulted in better performance for the MCF7 cell line compared to random initialization ensembles in terms of precision, it wasn't pursued further due to reliability issues of Q2 transcriptomic signatures. We observed many cases where the distance between Q1 signatures of compounds A and B was very small, e.g. 0.1, while the distance between signatures of compounds A and C, where C is a structural analogue of B (Tanimoto similarity > 0.85) and has a Q2 signature was high, e.g. 0.8. This kind of discrepancy between Q1 and Q2 signatures poses a problem for the learning model that only uses chemical structures as input.

## S4 Performance evaluation

**S4.1 Distance threshold for precision.** The model outputs a continuous value between 0 and 1 for the distance between compounds' affected biological processes (GO terms). In order to evaluate the model's precision, a reasonable distance threshold has to be specified. Compounds with predicted distances below this threshold are considered similar in terms of affected biological processes. First, the connection between the distance threshold and the average number of common GO terms in the most upregulated and downregulated GO terms, respectively, for all compound pairs in the dataset was examined and is presented in Figure S7. The average number of common GO terms decreases linearly as the distance threshold increases (Figure S7). Additionally, for each compound per cell line, the distance threshold equivalent of a 90% Connectivity score was calculated (Figure S8). For a specific compound X, a threshold equivalent of a 90% score indicates that only 10% of other Touchstone compounds have a distance from X smaller that this threshold. The 90% CMAP score is a widely accepted threshold to identify compounds with similar transcriptomic signatures. Finally, for MCF7, for which enough quality 1 duplicate compound signatures are available, the distribution of pairwise distances between duplicate signatures is presented in Figure S4 (black line). Based on the information provided in Supplementary figures 7, 8 and 4, a threshold of 0.2 was selected when evaluating the models' precision across all cell lines. When calculating the model's precision on test compounds that exhibit maximum structural similarity to all training compounds less than 0.3, this threshold was adjusted to 0.22, because in this case no samples had a predicted distance less than 0.2.

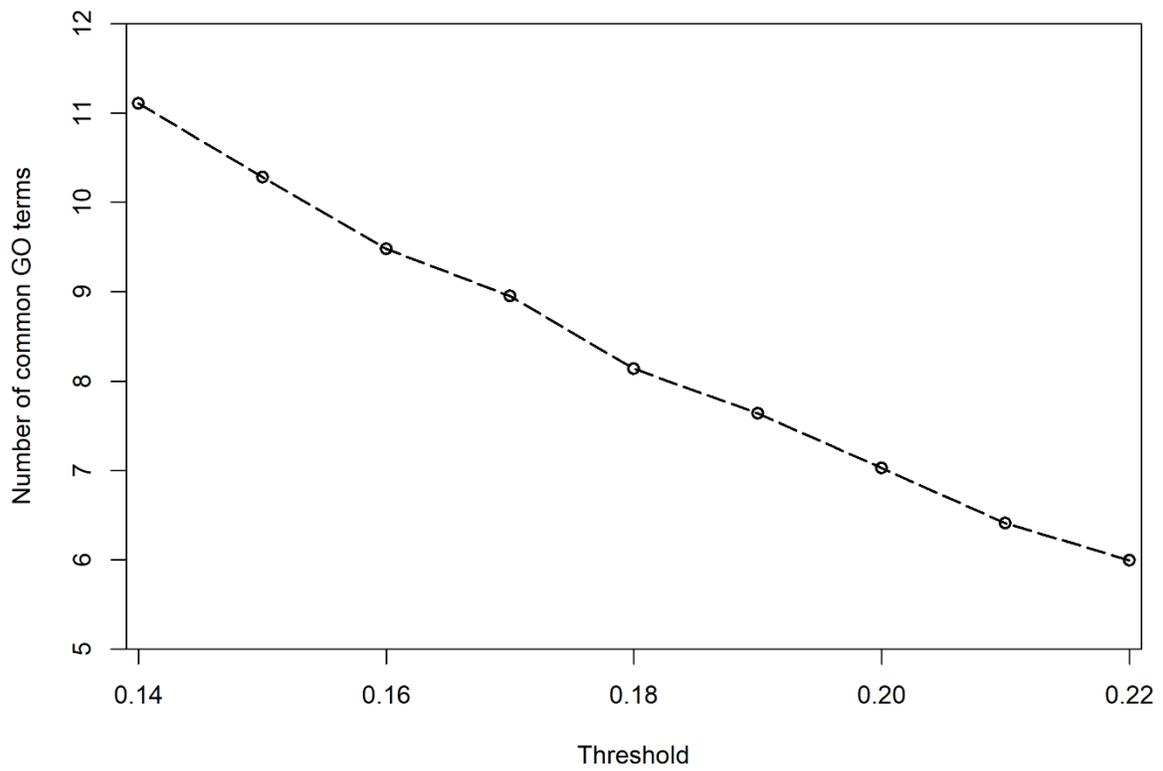

**Figure S7** The relationship between the biological effect distance threshold and the average number of common enriched BPs. In order to produce the above plot, signature pairs with GSEA distance below each threshold (x axis) are selected and the average number of common GO terms (BPs) in the 20 most upregulated and downregulated terms of all pairs is calculated (y axis).

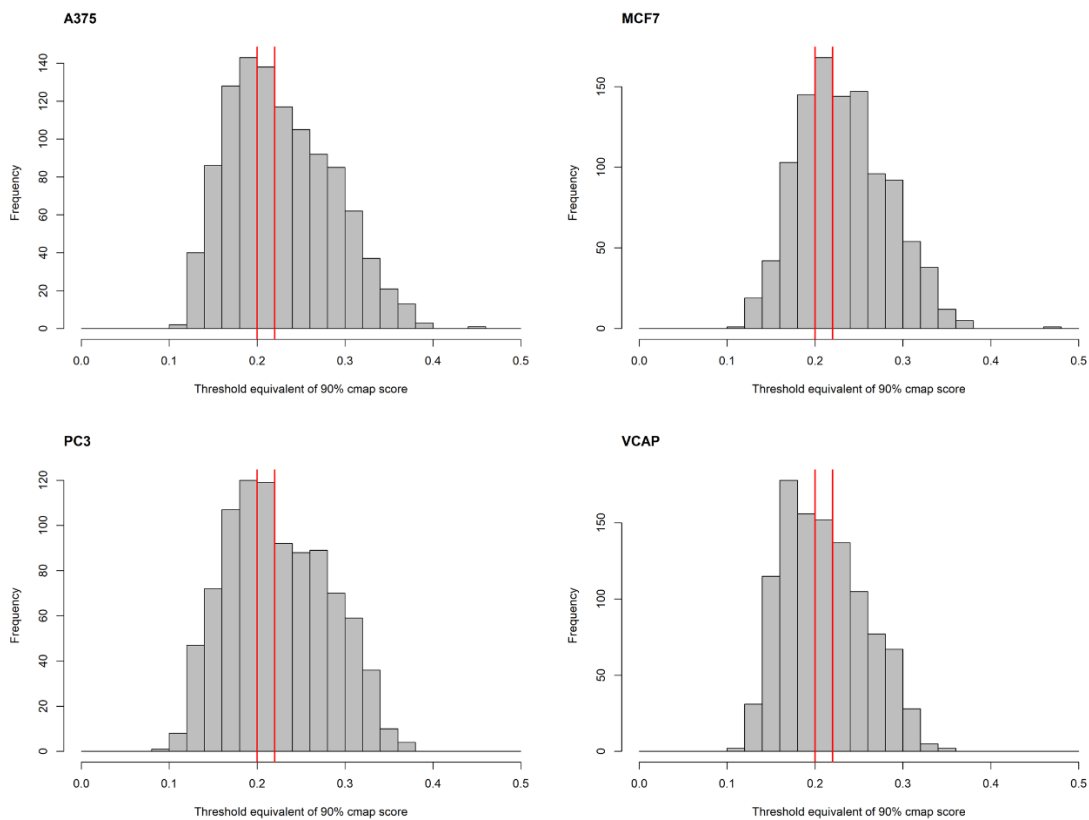

**Figure S8** Histogram of the threshold, which is equivalent to a 90% CMAP score, for all Touchstone compounds per cell line. The red vertical lines at 0.2 and 0.22 indicate the thresholds that were used to evaluate the models' precision. Across all Touchstone compounds, in all cell lines, the utilized thresholds are close to the mean of the threshold equivalent to a 90% CMAP score.

**S4.2 ReSimNet performance evaluation in all cell lines.** Our approach was compared with a recently proposed architecture called ReSimNet.[6] ReSimNet takes as input the 2048-bit ECFP4 fingerprints of two chemical compounds and predicts their CMap score, which corresponds to the transcriptional response similarity of their GEx signatures. ReSimNet encodes the ECFP4 input to embedding vectors in the latent space using Siamese MLPs and predicts their CMap score as the cosine similarity of their embeddings. In our implementation, ReSimNet was trained to predict the similarity between compounds' affected BPs and afterwards the output similarity was transformed to a distance value for evaluation. The performance of randomly initialized ReSimNet ensemble models was evaluated for each cell line, on its respective test set (Table S6).

**Table S6** Cell line specific performance of random initialization ReSimNet ensemble models

| Cell-line | MSE | MSE @1% | Pearson's r | Precision (%) | Precision top 1% | No.Predicted Similars |
|---|---|---|---|---|---|---|
| A375 | **0.012** | **0.022** | **0.6** | 32.23 | **56.80** | 18243 |
| VCAP | 0.039 | 0.105 | 0.38 | **32.69** | 52.97 | 9245 |
| PC3 | 0.017 | 0.032 | 0.49 | 25.02 | 46.89 | 14195 |

**S4.3 Cross validation performance.** For each cell line, we evaluated the performance of a 10 model ensemble in a 5-fold cross validation split. Each validation set contains pairwise distances between BPs of non-overlapping sets of 80 validation compounds and all remaining training compounds. When extracting validation compounds, the maximum allowed Tanimoto similarity between ECFP4 fingerprints of validation and training compounds was set to 0.85. The results of the 5-fold cross validation are presented in Table S7. In all tested cell lines, our approach was able to produce consistently good results.

**Table S7** Cross validation performance of deepSIBA

| Cell-line | MSE | MSE @1% | Pearson's r | Precision (%) |
|---|---|---|---|---|
| A375 | **0.008** | **0.005** | **0.56** | 92.11 |
| VCAP | 0.025 | **0.005** | 0.54 | 64.28 |
| PC3 | 0.011 | 0.009 | 0.54 | **96.47** |
| MCF7 | 0.013 | 0.009 | 0.52 | 58.94 |

## S5 Signaling pathway inference for target structure

**S5.1 Parameter selection.** The most important parameters of the signaling pathway inference are the distance threshold $d_{th}$ and the frequency threshold $f_{th}$. Training set compounds with predicted distances from the target less than $d_{th}$ are selected as its neighbors, while pathways that appear in the neighbors' signatures with frequency higher than $f_{th}$ are inferred as the target's signature. The performance of the inference method was evaluated for different values of $d_{th}$ and $f_{th}$ in the test set of MCF7 (Figure S9 and S10). The average precision of the inferred pathway signatures as well as their length were chosen as evaluation metrics. The performance of the method decreases as $d_{th}$ increases and $f_{th}$ is kept constant at 0.65 for both the upregulated and downregulated signatures. In terms of precision, as $d_{th}$ increases the precision of the approach decreases, and for $d_{th} > 0.4$ it becomes 0 (Figure S9A and S9B). In terms of the length (average number) of the inferred signatures, for $d_{th}$ higher than 0.4, the length of inferred signatures becomes 0, as more distant compounds are considered neighbors (Supplementary Figure 9C and 9D). After selecting 0.65 as a reasonable threshold for $d_{th}$, we evaluated the performance of the approach for different frequency thresholds $f_{th}$, the results are presented in Figure S10. As $f_{th}$ increases, while $d_{th}$ is kept constant at 0.2 the inference becomes more strict. This results in increased precision (Figure S10A and S10B), but shorter in length inferred pathway signatures (Figure S10C and S10D). Based on these results, the selected parameters of the pathway inference for the MCF7 cell line and the respective use case, are presented in Table S8.

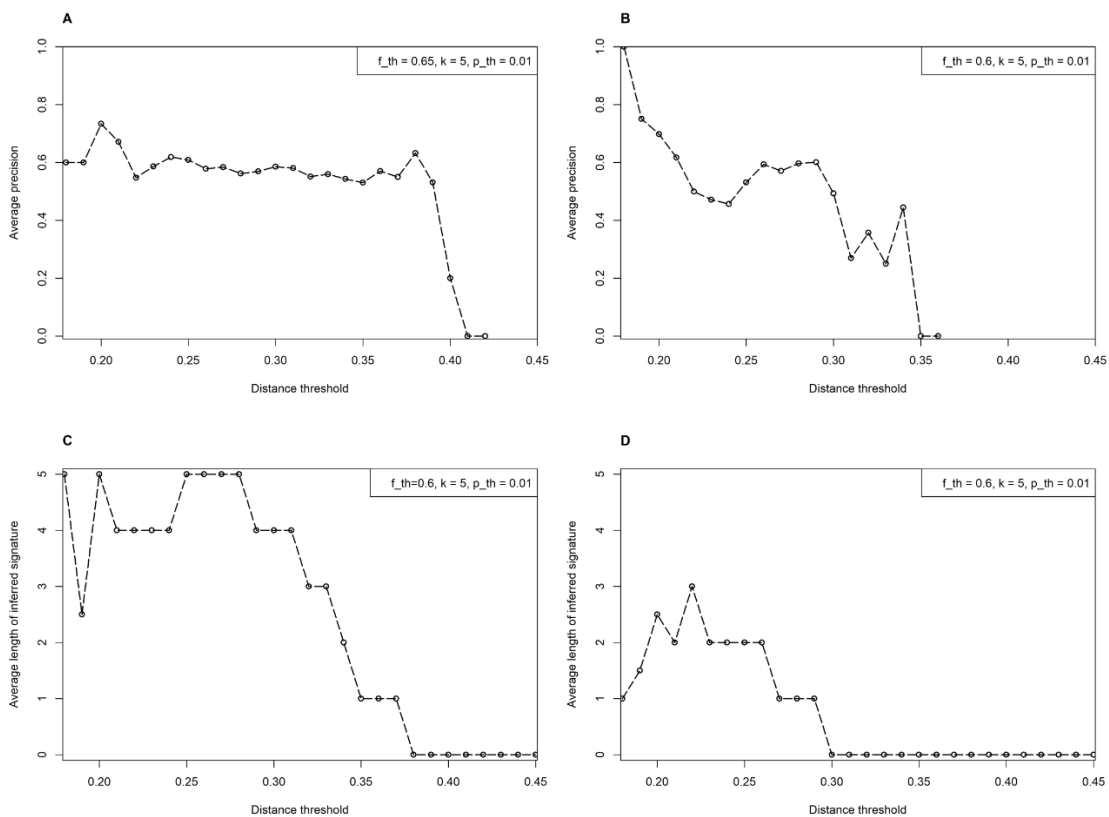

**Figure S9** Performance evaluation of the signaling pathway inference for different distance thresholds for the test compounds of the MCF7 cell line. Reference compounds (training) with predicted distance lower than the threshold are selected as the target's neighbors. The rest of the inference parameters are kept constant and their values are presented in the legend. (A) The average precision of the downregulated pathway signature as a function of the distance threshold; (B) The average precision of the upregulated pathway signature as a function of the distance threshold; (C) The average length of the inferred downregulated pathway signature; (D) The average length of the inferred upregulated pathway signature.

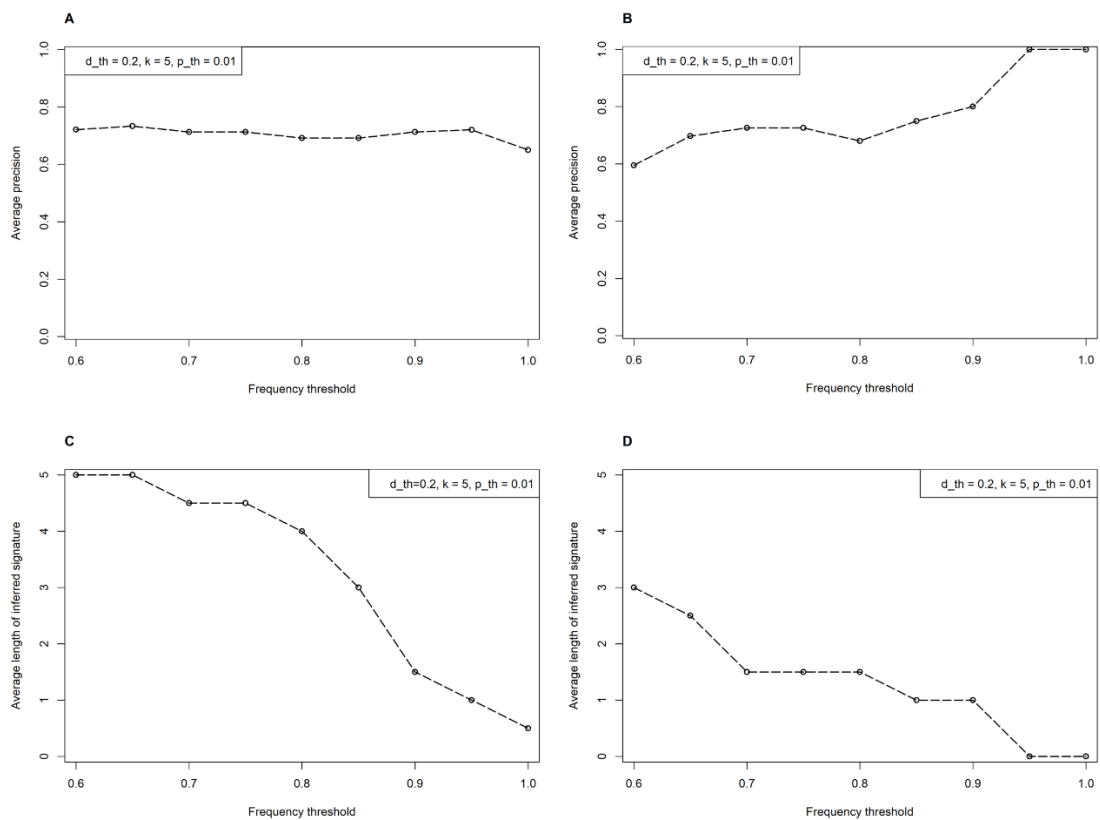

**Figure S10** Performance evaluation of the signaling pathway inference for different frequency thresholds for the test compounds of the MCF7 cell line. Signaling pathways that appear in the neighbors' signatures with frequency higher than $f_{th}$ are inferred as the target's signature. The rest of the inference parameters are kept constant and their values are presented in the legend. (A) The average precision of the downregulated pathway signature as a function of the distance threshold; (B) The average precision of the upregulated pathway signature as a function of the distance threshold; (C) The average length of the inferred downregulated pathway signature; (D) The average length of the inferred upregulated pathway signature.

**Table S8** Parameter values for the signaling pathway inference approach

| Parameter | Value |
|---|---|
| Distance threshold $d_{th}$ | 0.2 |
| Number of neighbors $k$ | 5 |
| Frequency threshold $f_{th}$ | 0.65 |
| P-value threshold $p_{th}$ | 0.01 |

## S6 References

1. RDKit: Open-source cheminformatics, http://www.rdkit.org/, (accessed 2020-03-20).
2. D. Rogers and M. Hahn, *Journal of chemical information and modeling*, 2010, **50**, 742-754.
3. F. Sirci, F. Napolitano, S. Pisonero-Vaquero, D. Carrella, D. L. Medina and D. di Bernardo, *NPJ systems biology and applications*, 2017, **3**, 1-12.
4. J. Lamb, E. D. Crawford, D. Peck, J. W. Modell, I. C. Blat, M. J. Wrobel, J. Lerner, J.-P. Brunet, A. Subramanian and K. N. Ross, *science*, 2006, **313**, 1929-1935.
5. D. K. Duvenaud, D. Maclaurin, J. Iparraguirre, R. Bombarell, T. Hirzel, A. Aspuru-Guzik and R. P. Adams, *Journal*, 2015, 2224-2232.
6. M. Jeon, D. Park, J. Lee, H. Jeon, M. Ko, S. Kim, Y. Choi, A.-C. Tan and J. Kang, *Bioinformatics*, 2019, **35**, 5249-5256.